\documentclass{eas}
\usepackage{graphicx}
\begin{document}

\title{A new chemodynamical tool to study the evolution of galaxies in the
  local Universe} 
\runningtitle{Champavert \& Wozniak: A new chemodynamical tool\dots}

\author{Nicolas Champavert}
\address{Universit\'e Lyon 1; CRAL, Observatoire de Lyon, 9 avenue Charles
  Andr\'e, Saint-Genis Laval cedex, F-69561, France;
  \email{champavert@obs.univ-lyon1.fr}
}
\author{Herv\'e Wozniak}
\sameaddress{1}
\begin{abstract}
We present some preliminary results obtained with a new galactic
chemodynamical tool under development.  In the framework of
non-instantaneous recycling approach, we follow the interactions due
to star formation and feedback processes.
One of the main original features of our code is that we record the
abundance evolution of several chemical elements.  This allows us to
build cooling functions dependent on the real abundances of individual
elements.  We illustrate the need for such metal-dependent cooling
functions using a toy model made of a star cluster embedded in a
two-phase gas cloud.  Our results suggest that computing cooling rates
according to individual abundances of chemical elements can influence
the star formation rate.
\end{abstract}
\maketitle
\section{Introduction}
Dynamical and chemical evolution of galaxies are closely related.  The
complex spatial distribution of metals in galaxies reflect 
strong ties.  Indeed, heavy elements, synthesized by successive
generations of stars, enrich the interstellar medium (ISM) through
stellar winds and supernova explosions. Density waves, shear,
gravitational perturbations are then responsible to mix the ISM and
dissolve young stellar clusters.  Furthermore the metal enrichment
affects the thermal evolution of the gas because cooling rates are
very sensitive to the chemical composition.  Besides mass feedback,
stars are also responsible for energy feedback which both influences
the dynamics and the temperature of the gas.  Therefore we need to
compute chemical and dynamical evolution self-consistently.

\begin{figure}[htbp]
\centering
\resizebox{!}{8cm}{\includegraphics{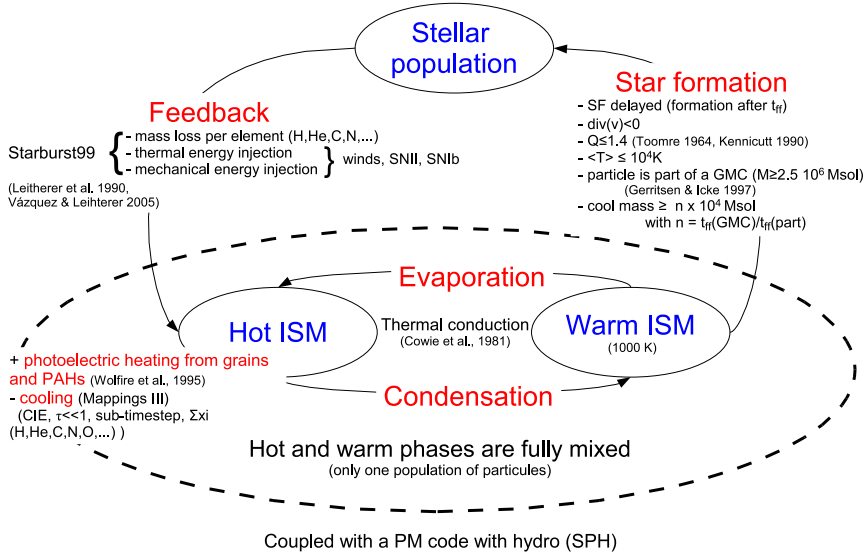}}
\caption{Schematic presentation of our new chemodynamical tool}
\label{schema}
\end{figure}

Figure~\ref{schema} shows the different physical interactions between
the stars and the ISM handled with our tool. It is designed to be
coupled with an hybrid particle-mesh N$-$body $+$ smooth particle
hydrodynamics (SPH) code to compute high-resolution chemodynamical
simulations of galaxies.  We will describe the two components (star
and gas) and the different aspects of the interplay between them.

\section{Evolution of stellar populations}
\label{stellar_population}
In our simulations, stars are in fact collisionless particles of
masses comparable to typical stellar clusters. The mass and energy released
from stellar particles are distributed to the neighbouring gas
particles.  Energy feedback from both the stellar winds and the
supernovae is considered.  The time evolution of all relevant
quantities is computed with the evolutionary synthesis code
Starburst99~v5.1 (Leitherer et al.~\cite{leitherer99}, V\'azquez \&\ Leitherer~\cite{vazquez05}).  The choice of
Starburst99 was motivated by the possibility to obtain the time
evolution of mass lost by a single stellar population (SSP) for several individual chemical
elements, namely H, He, C, N, O, Mg, Si, S and Fe. Starburst99 is, at our
knowledge, the only public code to provide them. Moreover, it offers a
large choice of inputs and outputs.  We can choose, in particular, the
timescale of star formation (instantaneous in our case), the IMF (a
multi-power-law), the metallicity of the SSP (5
metallicities are available from 0.02 to 2.5~solar metallicity), the
stellar evolutionary tracks (Padova or Geneva). We can obtain the time
evolution of SNe rates, the energy losses from stellar winds and SNe,
the mass losses for several chemical elements, the number of ionizing
photons, the SSP spectra. Only SNII and SNIb losses are provided by Starburst99
but feedback from SNIa will be soon implemented
directly in our code with different recipes.

Let us consider a stellar population characterized by a Kroupa IMF
with exponents (1.3, 2.3) and mass boundaries (0.1, 0.5, 100\,M$_\odot$).
We consider all the gas released by the stellar population (stellar winds and
SNe) from birth to 5\,Gyr.
$m_X$ is the mass fraction of the element $X$ among all the metals
i.e. the ratio between the mass of the element $X$~($M_X$) and
the total mass of metals ($M_Z$). $m_{X(Z_\odot)}$ is the same ratio
but for a gas with solar metallicity.
Typical mass fraction at solar metallicity are
H~($7.06\ 10^{-1}$), He~($2.75\ 10^{-1}$), C~($3.03\ 10^{-3}$),
N~($1.11\ 10^{-3}$), O~($9.59\ 10^{-3}$), Mg~($5.15\ 10^{-4}$),
Si~($6.53\ 10^{-4}$), S~($3.96\ 10^{-4}$) and Fe~($1.17\ 10^{-3}$).

\begin{figure}[t]
\begin{center}
\resizebox{!}{5cm}{\includegraphics{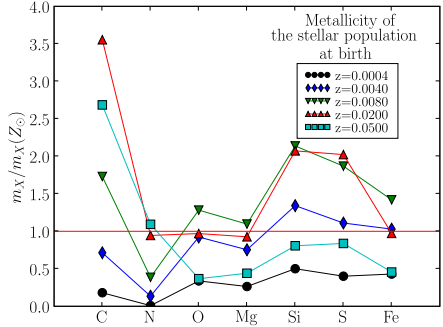}}
\caption{Ratios of the mass fraction of elements in the ejected gas ($m_X$) and of the
  mass fraction of the same elements in gas with solar chemical composition
  ($m_X(Z_\odot)$). We consider all the gas ejected from birth to
  5\,Gyr. These ratios are shown for several chemical elements and for SSP of different initial metallicities.}
\label{abundances}
\end{center}
\end{figure}

In Fig.~\ref{abundances}, we
display the ratio $m_X/m_{X(Z_\odot)}$ in the released gas for several
elements and for different initial metallicities of the stellar population.
By definition, if the abundance ratios of metals in the released gas are the same as for 
solar composition, $m_X/m_{X(Z_\odot)}$ should be equal to 1.
However, the metal abundance ratios are not solar in the released gas. 
In particular, for high metallicities the released gas is particularly enriched in 
carbon, silicon and sulfur.
This gas will be mixed with the rest of the ISM resulting in a gas
with non-solar abundance ratios.

\section{Interstellar medium description}
The ISM is approximated by a two-phase model: an isothermal warm phase~($10^3$\,K)
and a hot phase whose temperature is allowed
to vary between 10$^4$\,K and 10$^8$\,K.
We assume that the two phases are fully mixed because the typical size of SPH particles (several parsecs) does not allow to resolve them. 
Thus, each SPH particle contains both a warm and a hot phase.
The warm ISM mass fraction can increase (condensation) or decrease (evaporation) due to
thermal conduction (Cowie et al.~\cite{cowie81}). It also obviously decreases when star formation
operates because star particles are created in the warm phase.
The hot phase is heated by the photoelectric heating from grains and PAHs (Wolfire et al.~\cite{wolfire95}) 
and by stellar feedback due to stellar winds and
supernovae. 
We decide to inject all the released mass and energy from stars into the {\em hot}
phase of the gaseous neighbouring particles. The redistribution of energy
between the two phases is achieved at the next timestep thanks to all physical
processes described above.
Moreover, a fraction of the stellar energy feedback can be converted to mechanical energy that can affect directly the velocity field of
the surrounding gas.

\begin{figure}[t]
  \centering
  \resizebox{6cm}{!}{\includegraphics{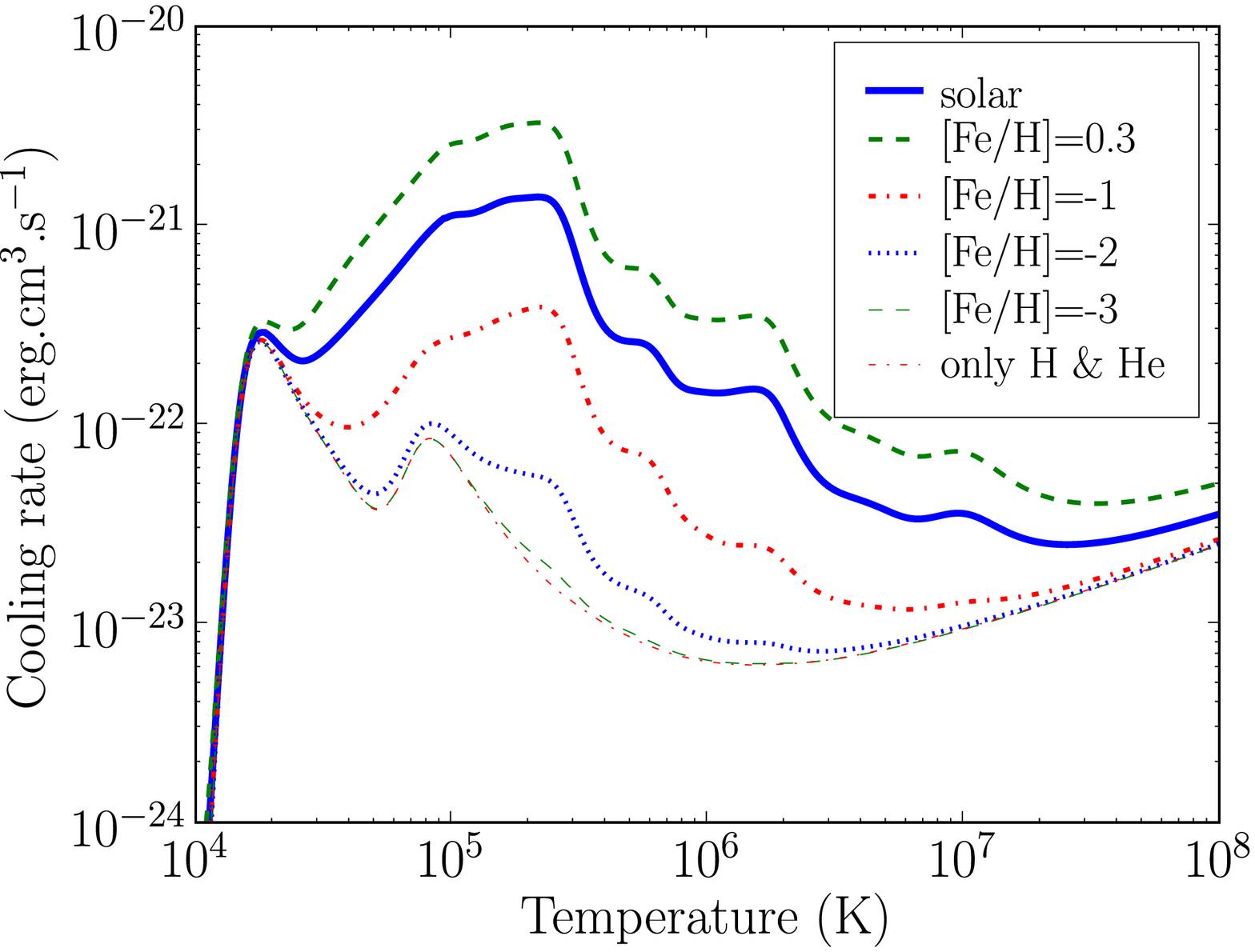}}
  \resizebox{6cm}{!}{\includegraphics{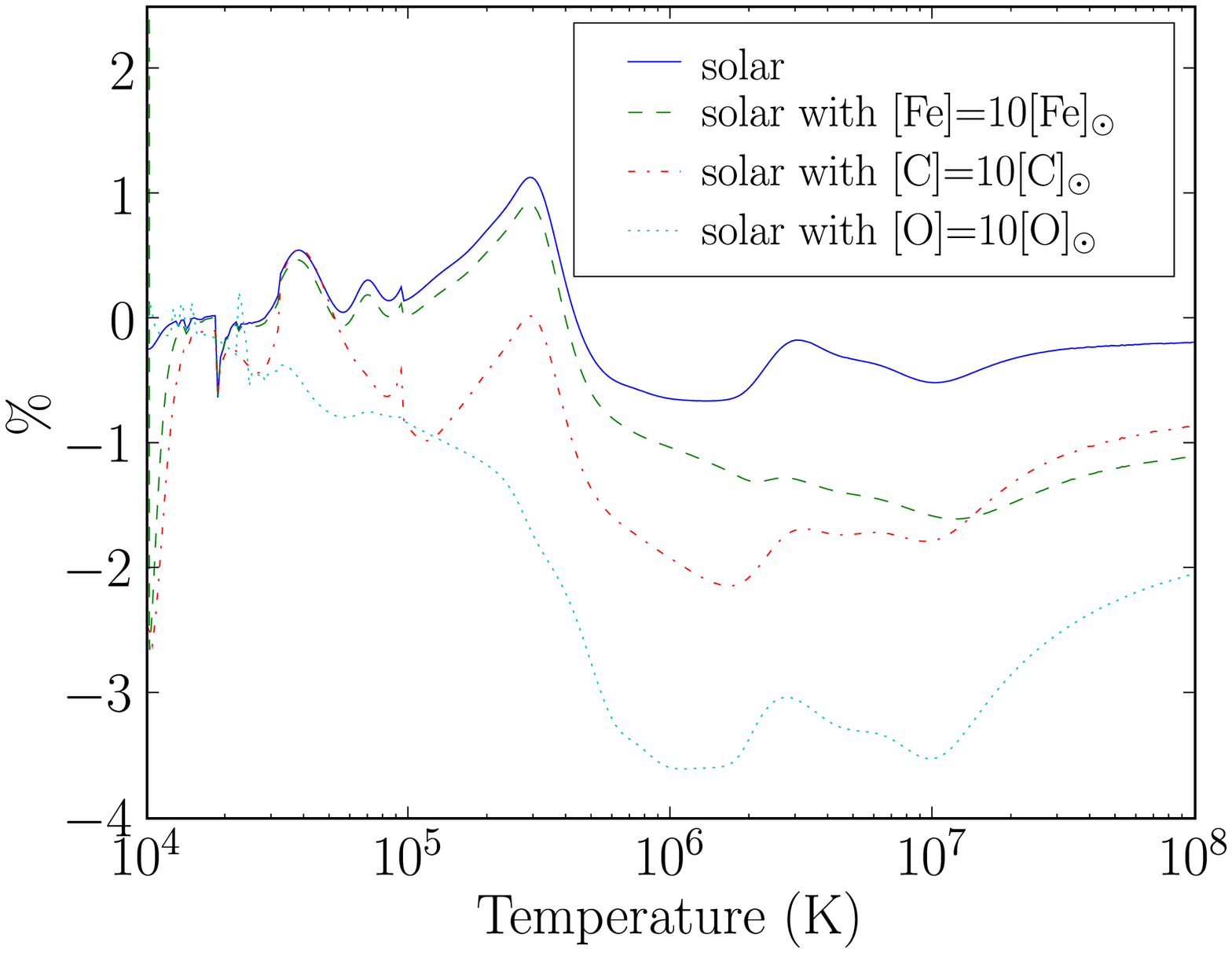}}
  \caption{\textbf{Left panel:} Cooling rates as a function of temperature computed with 	 
    Mappings~III for different metallicities. The abundances used in
    calculations are taken from Sutherland \& Dopita (\cite{s&d93}).
    \textbf{Right panel:} Percentage of relative errors made when
    reconstructing cooling function with our recipe for different chemical compositions.
  }
  \label{cooling}
\end{figure}
It is well known that cooling rates are sensitive to the chemical composition of the gas.
At solar composition the cooling rate is greater by more than an
order of magnitude for some temperatures than for a gas without metals
(Fig.~\ref{cooling}, left panel).
As for SSP evolution, standard cooling functions used in chemodynamical studies are generally computed for different
metallicities but keeping solar abundance ratios (e.g. Boehringer \& Hensler, \cite{b&h89}) or
with solar abundance ratios and enhanced abundances for some specific elements (e.g. Sutherland \& Dopita, 
\cite{s&d93}).
Hence we need to take into account the real abundances of each element in
the gas to build more realistic cooling functions whichever the chemical
composition.
We thus consider the hot phase of the ISM as an optically thin gas in collisional
ionization equilibrium.
Cooling calculations are performed in the temperature range $10^4$\,K to
$10^8$\,K. Computing cooling rates with Mappings~III 
for each gas particle at each timestep of a chemodynamical simulation is a very CPU time
consuming task. Therefore we use a recipe to reconstruct cooling rates on the fly
from individual abundances of elements (see Champavert \& Wozniak, \cite{cw07}, for a full description). In short, this recipe comes down to
a linear combination of individual cooling curves precomputed with Mappings~III.
The relative errors between our reconstructed cooling curves and those
calculated with Mappings~III remain below a few percent
(Fig.~\ref{cooling}, right panel). Thus, they are comparable to others coming from, for instance, the hydrodynamical scheme.

\section{Star formation recipes}
We assume that star formation occurs if the following criteria are satisfied.
The region must be contracting ($\textrm{div}(v)<0$). It has to be unstable
according to the Toomre's instability criterion (Toomre, \cite{toomre64}): $Q \leq
\lambda$. We take $\lambda = 1.4$ derived
from observation by Kennicutt (\cite{kennicutt90}). The gas must be cold enough:
\mbox{$T<10^4$\,K} with $T$ the mean gas temperature of a SPH particle. This threshold has been chosen because of 
our cooling functions which are only available for $T\geq 10^4$~K for
the moment. Finally the gas particle must belong to a giant molecular cloud
(Gerritsen \& Icke, \cite{g&i97}).
A gas particle is assumed to be part of a giant molecular cloud (GMC) if the
mass of the cloud containing this particle, i.e. the mass of this particle
with its gaseous neighbours, is larger than $2.5\,10^6$\,M$_\odot$, typical
mass of a big GMC.
When a gas particle fulfills all these conditions, it becomes eligible for star formation. However, a stellar particle is formed only 
after a delay equal to a free-fall time (t$_{\mbox{ff}}(\mbox{part})$) to mimic the gravitational collapse of the molecular cloud. 
The mass of the new stellar particle is
set to $n\times 10^4$\,M$_\odot$ where
$n = t_{\mbox{ff}}(\mbox{GMC})/t_{\mbox{ff}}(\mbox{part})$ is the number of elemental stellar clusters formed in the GMC.
We obviously need a warm phase more massive than $n\times 10^4$\,M$_\odot$
because the mass needed to form stars is deducted only from the warm phase.
Each new stellar particle inherits the chemical composition from the gas in which 
it was born.

However, the evolution of a SSP, as predicted by Starburst99, only
depends on the global metallicity $Z$ because stellar evolutionary
tracks are generally computed with solar abundance ratios or with a
particular enhancement for some elements which is unavoidably
different from what we obtain with our code.  In order to be able to 
self-consistently compute the chemical evolution of individual elements in the future, we will obviously need
data from evolutionary tracks including variations on the abundances
of individual elements.  At the moment, when a gas particle should form a new stellar particle, we have to assign
the global metallicity of the gas to the new particle.  For a given metallicity and IMF, we compute linear
interpolations between values pre-calculated with Starburst99 to obtain
the mass and energy losses of the SSP.
\section{A test model}
Let us now illustrate one of the main properties of our tool.
We consider a gas particle of 10$^7$\,M$_\odot$ with 30\% of mass in the warm
phase. The density of the hot phase is 0.005\,cm$^{-3}$. At $t=0$, the gas is
made of hydrogen~(77\%) and helium~(23\%), its temperature is 10$^4$\,K
and a stellar particle of 1.4 10$^5$\,M$_\odot$ is born.
The star formation delay is constant (10\,Myr) and all the subsequent stellar particles
formed have the same mass (1.4~10$^5$~M$_\odot$). All other parameters are 
described in Sect.~\ref{stellar_population}.
We have calculated the evolution for 2 models. In model~A, the cooling rate depends only on $Z$ with 
solar abundance ratios. In model~B, cooling rates are obtained according to the real
abundances of chemical elements.

  \begin{figure}[t]
\resizebox{4.1cm}{!}{\includegraphics{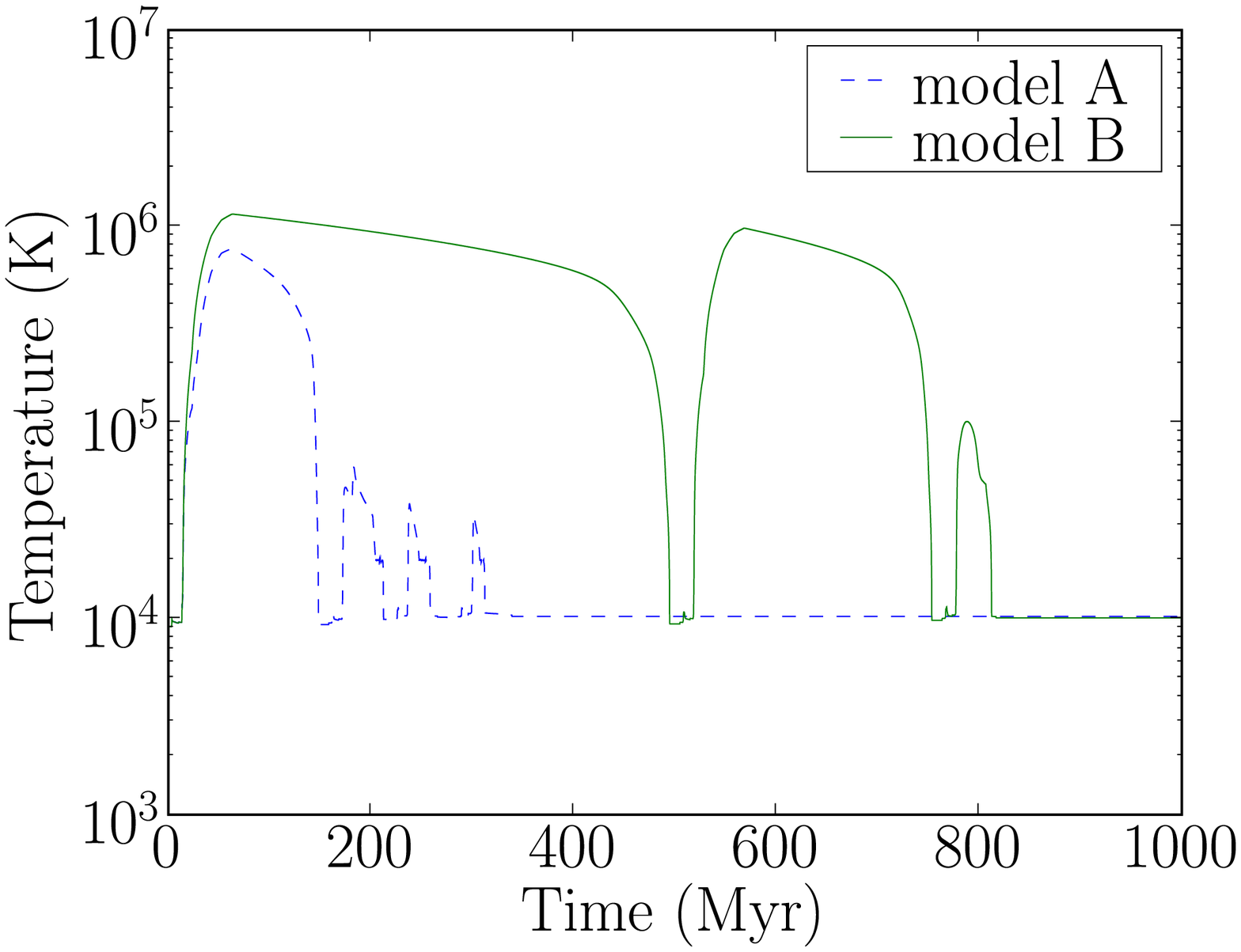}}
\resizebox{4.1cm}{!}{\includegraphics{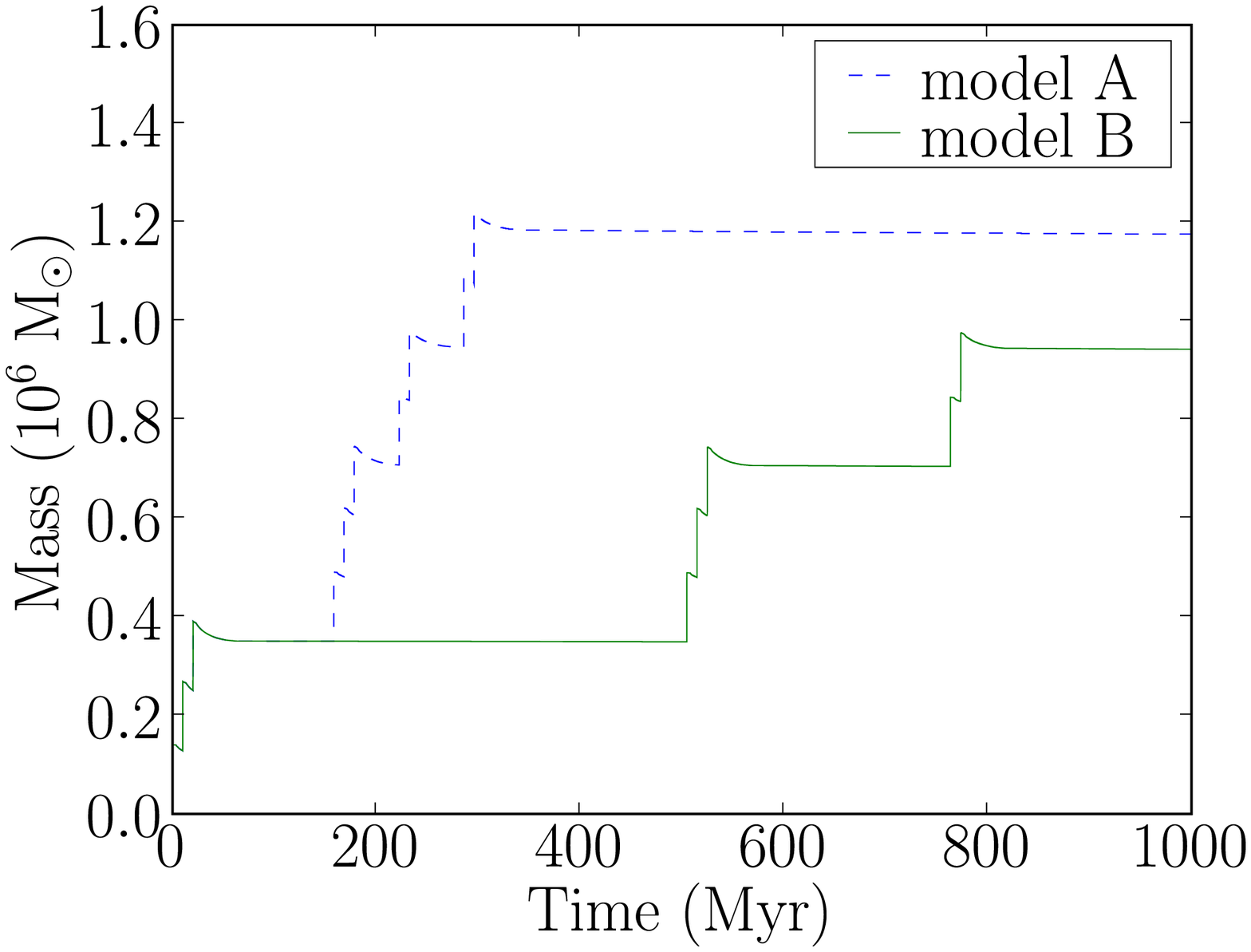}}
\resizebox{4.1cm}{!}{\includegraphics{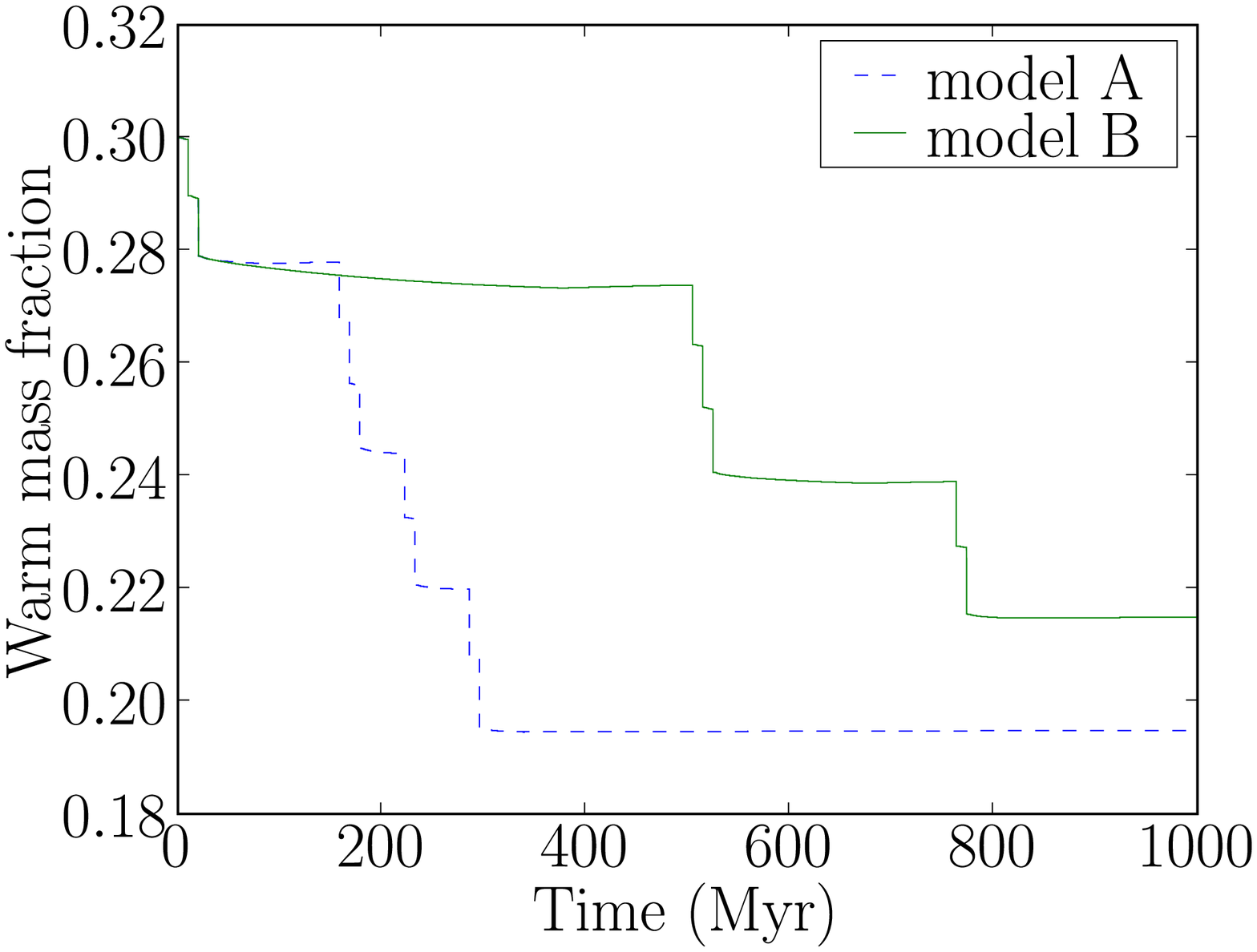}}
    \caption{Evolution of the mean gas temperature (\textbf{left panel}),
    the stellar mass (\textbf{middle panel}) and the warm mass fraction
    (\textbf{right panel}) for the 2 models.
    Solid line is for model~A (only global metallicity is calculated) and
    dashed line is for model~B (abundances for each chemical element are calculated).}
    \label{evol}
  \end{figure}
Figure~\ref{evol} shows the evolution of the mean temperature of
the warm and hot phase (left panel), of the stellar mass
(middle panel) and of the warm mass fraction~(right panel) for both models.

Cooling is clearly less efficient when we handle the abundances of elements
leading to an higher mean gas temperature. This therefore leads to a lower star formation rate since the mean temperature remains
for a longer time above the temperature threshold for star formation ($10^4$\,K for our models). 
At $t = 1$\,Gyr, the cumulated stellar mass formed in 
model~A is about 1.25 times greater that for model~B.
As this longer timescale alters the star formation history, the warm gas mass fraction also evolves differently.


\end{document}